# The Scaling Relationship between Citation-Based Performance and Scientific Collaboration in Natural Sciences

**Guillermo Armando Ronda-Pupo**

Departamento de Administración
Universidad Católica del Norte, Chile
Universidad de Holguín, Cuba
gronda@ucn.cl

**J. Sylvan Katz**

SPRU, Jubilee Building
University of Sussex
Falmer, Brighton, BN1 9SL, UK

Johnson-Shoyama Graduate School of Public Policy,
University of Saskatchewan Campus,
Diefenbaker Building, 101 Diefenbaker Place,
Saskatoon, SK, Canada, S7N 5B8

j.s.katz@sussex.ac.uk

## Abstract

*The aim of this paper is to extend our knowledge about the power-law relationship between citation-based performance and collaboration patterns for papers in the natural sciences. We analyzed 829,924 articles that received 16,490,346 citations. The number of articles published through collaboration account for 89%. The citation-based performance and collaboration patterns exhibit a power-law correlation with a scaling exponent of $1.20 \pm 0.07$. Citations to a subfield's research articles tended to increase $2^{1.20}$ or 2.30 times each time it doubles the number of collaborative papers. The scaling exponent for the power-law relationship for single-authored papers was $0.85 \pm 0.11$. The citations to a subfield's single-authored research articles increased $2^{0.85}$ or 1.89 times each time the research area doubles the number of non-collaborative papers. The Matthew effect is stronger for collaborated papers than for single-authored. In fact, with a scaling exponent < 1.0 the impact of single-author papers exhibits a cumulative disadvantage or inverse Matthew effect.*

# Introduction

Collaboration has been a constitutive aspect of science from its very beginning, as it was one way to transmit and improve knowledge (Archambault, Beauchesne, Côté, & Roberge, 2014). Scientists from several academic disciplines have studied scientific collaboration in the past 25 years. For example, in the Web of Science *category Information Science & Library Science* appear 7,911 pages in 591 articles published in 71 journals to disseminate findings about scientific collaboration. The journal *Scientometrics* accounts for 34% of the overall scientific output about this subject matter. 57% of the papers come from the last seven years suggesting exponential growth.

Among the most influential papers on the scientific collaboration topic are Katz and Martin (1997), Katz (1994), Persson, Glanzel, and Danell (2004), Beaver (2001), Luukkonen, Tijssen, Persson, and Sivertsen (1993), and Hara, Solomon, Kim, and Sonnenwald (2003). These five papers account for the 42% of the overall impact of this line of research in the last 25 years and they contributed fostering academic interest towards the study of collaboration.

In the past few years, the academic debate of research on scientific collaboration has turned into a discussion about the effect that collaboration has on the impact of research papers. The trend of studying the influence of collaboration on the impact of articles has been widely supported (Katz & Hicks, 1997; Katz & Martin, 1997; Kliegl & Bates, 2010; Tang & Shapira, 2010; Zhai, Yan, Shibchurn, & Song, 2014). Recently, Avkiran (1997); Elena Luna-Morales (2012); Glänzel (2002); González-Teruel, González-Alcaide, Barrios, and Abad-García (2015); Rousseau (2000); Rousseau and Ding (2015); Van Raan (1998) ). These authors have published findings about the existence or absence of a relationship between collaboration and the impact of articles in several scientific fields.

Katz (2000, p. 35) suggested that studying the power-law or scale-independent relationship between the impact of co-authored papers and group sizes would facilitate a better understanding of the impact of a research field's collaborative research activity within and across science systems. A system with a scale-independent property statistically exhibits that property at many levels of observation and it is mathematically described by power-law distributions and correlations (Katz, 2006a). Co-authorship networks that emerge out of the process of creating and disseminating knowledge represent the collaboration structure. These networks tend to have in-link vertex connectivity that follows a power-law distribution (Barabasi & Reka, 1999).

A few studies have analyzed the power-law relationship between collaboration and impact. Archambault et al. (2014) found that collaboration intensity follows a power-law, and that the larger an entity, the less it tends to collaborate intensely with outside partners. Katz (2000) suggested that one could derive performance indicators by dividing observed values by expected values calculated using a power law regression on the data addressing the non-linear properties of collaboration (Archambault et al., 2014).

Recently, Ronda-Pupo and Katz (2015) found a power-law relationship between citation-based performance and collaboration for articles in journals in the field of management. The relationship between citation-based performance and collaboration patterns in fields of natural sciences has not been done. The aim of this paper is to explore the power-law relationship between citation-based performance and scientific collaboration for these fields.

## Theory and hypothesis

de Solla-Price (1963) first suggested the existence of a power-law citation distribution for the publishing activity of scientists. Clauset, Shalizi, and Newman (2009, Table 6.1) found that citations to papers had a good likelihood of being a power-law distribution. Recently, Brzezinski (2015) found power-law distribution for highly cited papers of the disciplines Physics and Astronomy on the Scopus database. This author concluded that the power-law hypothesis is a plausible one for around half of the Scopus fields of science. We pose as the first hypothesis:

> *H1: the distribution of the citations to overall/ collaborative/ single-authored peer-reviewed articles in the natural sciences will follow a power law distribution.*

A positive and significant power-law correlation between international collaboration and size of countries was found to have an exponent of 1.14± 0.03 ($R^2 = 0.95$) by Katz (2000).

Recently, Ronda-Pupo and Katz (2015) reported the existence of a power-law correlation between citations and collaboration with scaling parameter of $1.89 \pm 0.08$ in the articles in management journals. Based on these results, we expect a power-law correlation between citations and collaboration may exist across the subfields of natural sciences. We pose as second hypothesis:

> *H2: The citations to collaborated papers will show a power-law correlation with a scaling factor α > 1.*

Lotka (1926) first studied the distribution of scientific productivity. He found that a few authors accounts for approximately the 80% of the overall scientific output. de Solla-Price (1963) defined it as the Lotka's Law. A scientist already rewarded for their achievements get a higher chance of being rewarded once again, so that they become a part of an elite group enjoying preferential access to scientific resources and facilities (de Bellis, 2009). Merton (1968, 1988) called it success-breeds-success phenomenon by which the rich get richer while the poor get comparatively poorer the "Mathew Effect," after a well-known verse in the Gospel according Mathew (Mathew 25:29, King James version).

The exponent of a power-law correlation is a measure of the Mathew Effect of the citation-based impact. Ronda-Pupo and Katz (2015) found that the Mathew Effect of the citation-based performance of articles in the journals of the field of management is bigger for collaborative papers. The exponent for collaborative articles is 1.89 ±0.08 while for single-authored articles the exponent of 1.35 ±0.08. We expect the Mathew Effect will be bigger for the citation-based performance of collaborative articles across the fields of science given earlier. We pose as third hypothesis:

> *H3: The Mathew Effect for citations to collaborated papers will be bigger than to single-authored papers.*

**Methods**

The methodology used in the study comprises two main steps. First, we test the hypothesis for the power-law distribution on the citations to overall/ collaborated/ single-authored papers. Second, we run a power-law regression on numbers of citations and numbers of collaborative/non-collaborative papers.

*Verification the power-law distribution*

In this step we used the Clauset et al. (2009, Box 1) three-step procedure.

*Estimate the $x_{min}$ and scaling parameter α of the power-law*

The objective of this step is to determine the $x_{min}$ and the scaling parameter α. The $x_{min}$ is the value when the power-law begins or the lower bound on the scaling region. That is, the $x_{min}$ value is the highest probability point in the distribution where the power-law begins. To calculate the $x_{min}$ value we used the Formula 3.7 by Clauset et al. (2009) because the data is discrete. We used the computational program by Gillespie (2015). The scaling parameter α is calculated through the method of maximum likelihood.

*Calculate the goodness of fit between the data and the power-law*

The objective of this step is to find out if the hypothesis of the power-law distribution according to the data is a plausible one. For this step, we ran 2,500 Monte Carlo simulations. To determine the required number of samples to run for an accuracy of two decimal digits (for a two decimal digits $\varepsilon = 0.01$) we used the formula proposed by Clauset et al. (2009) $1/4\varepsilon^{-2}$. The (Clauset et al., 2009) formula suggests to run 2,500 samples to test if the distribution really follows a power-law.

We fit each sample individually to its own power-law model and calculated the *Kolgomorov-Smirnov (KS)* statistic for each one relative to its own model. The *p* value is the fraction when the resulting statistic is larger than the value for the empirical data (Clauset et al., 2009). We ran a goodness of fit test, which generates a *p* value that quantifies the plausibility of the hypothesis. According to Clauset et al. (2009) if the value of *p* is large (close to 1) then, the hypothesis that the data follows a power-law distribution is correct. We made the choice that the power-law is ruled out if $p \leq 0.10$, following Clauset et al. (2009).

*Compare the power-law with alternative hypothesis*

It is important to note as Clauset et al. (2009) suggest that a large *p*-value does not necessarily mean that the data obey to a power-law distribution. Thus, the objective of this step is to verify if the power-law distribution is the better fit to the data under analysis. For this step, we compare the power-law distribution to log normal, exponential and power-law with exponential cut-off as competing distributions using the *KS* test to measure the distance between distributions.

We calculated the *p*-value for a fit of each of the competing distributions and we compare it with the *p*-value of the power-law distribution. Finally, we used the likelihood ratio test for each alternative under comparison to make a decision if the data follows a power-law distribution or not. If the likelihood ratio is significantly different from zero the sign indicates, whether the alternative is favored over the power law model or not (Clauset et al., 2009). If the sign is positive, the power-law is favored over the alternative. And the distribution with most negative likelihood ratio and significant p-value is a better fit than the power law distribution.

**Power-law correlation between citation-based performance and collaborative activity**

**The model**

The model for the study follows the power-law approach,

$$CBP = kc^n \text{ (equation 1)}$$

Where *CBP* stands for citation-based performance, *c* for number of collaborative papers, *k* is a constant (intercept) and *n* is the scaling factor (slope of the log-log regression line).

*Definition of variables in the model*

Table 1 shows the variables and their conceptual definitions.

**Table 1**. Variables and their conceptual definitions.

| Variable | Conceptual definition |
|---|---|
| Citation-based performance | is the number of citations to receive by each paper within a field/subfield. |
| Collaboration | is the number of articles published with the participation of more than one author. |
| No collaboration | is the number of articles published with the participation of one author. If the author signs by more than one institution, was considered no-collaboration. |

*Data source*

The data for the study consists of publications in natural sciences published in the WOS database between 2005 and 2007, inclusive. We used the following publication types: articles including proceeding papers published in journals, letters, notes and reviews. We used these publication types for two reasons: 1) they are peer-reviewed and 2) they are a primary route for disseminating new knowledge in most scientific disciplines (Adams & Gurney, 2013).

To retrieve the data we used the tag Advance search SO= 'Journal Name' Refined by: Document Types: (Article OR Review OR Letter OR Note OR Proceedings Paper). We filter the document types Review and Proceedings paper by the ISI field PJ to ensure that the review is not a book review and the proceeding is a journal paper. Timespan: 2005-2007. Indexes: SCI-EXPANDED, SSCI, A&HCI. We retrieved the data by journals to avoid duplicates because the WOS categories allow assigning documents to more than one category. To download the records we added the results to marked list, next we saved the records by 500 (Because of WOS constrains) to Tab-delimited. Then we created an Excel database for the quantitative analysis. To assign each journal/paper to a unique field/subfield we used the Science Metrix journal classification[1]. The Science Metrix classification is similar to the NSF journal classification and it was updated recently.

---

[1] Available from Science Metrix web site http://science-metrix.com/en/news/science-metrix-launches-the-second-public-release-of-its-multilingual-journal-classification.

The Science Metrix classification scheme assigns journals to one of 176 subfields that can then be uniquely aggregated into 22 fields. A problem with this classification is that prestigious multidisciplinary journals like Science, Nature, *Plos One*, etc. are assigned to single subfields of General Science and Technology because of the difficulty of assigning individuals papers to unique subfields. The articles in multidisciplinary journals tend to receive more citations than articles in subfield focused journals. For this analysis, we examined 33 natural science subfields in the fields of (1) biology, (2) chemistry, (3) earth & environmental sciences, (4) mathematics & statistics and (5) physics & astronomy.

## Results

Table 2 shows the results of the Citation-Based Performance according to collaboration patters. The scientific community of the fields in the natural sciences published 825,922 articles in 1,898 journals between 2005 and 2007, inclusive.

**Table 2**. Summary of citation based performance according to collaboration patterns.

| Collaboration patterns | Nº Papers | % Total | Citations | % Total |
|---|---:|---:|---:|---:|
| Collaboration | 726,306 | 88% | 15,257,054 | 93% |
| No Collaboration | 99,616 | 12% | 1,233,292 | 7% |
| Total | 825,922 | 100% | 16,490,346 | 100% |

The main patterns found could be summarized in the following items:

- All 5 fields and 33 sub-fields published more co-authored papers than single-authored.
- The co-authored papers received more citations than single-authored papers.
- The co-authored papers account for the 93% of the citations. It suggests that collaborated papers receives 16 times the number of citations to single-authored papers.
- The median citations of collaborated papers is 10 while for single-authored it is 4.

### *Verification of the power-law distribution*

### *Estimating the $x_{min}$ and scaling parameter α of the power-law*

We tested the fit of three data sets to power-law distributions: the citations to all papers, the citations to collaborative papers and the citations to single-authored papers. Table 3 shows the results of fitting the data to a power-law distribution to 2,500 iterations through Monte Carlo bootstrapping analysis.

**Table 3**. Results of fitting the power-law to the datasets.

| Dataset | $X_{min}$ | α | p | KS |
|---|---|---|---|---|
| Overall | 22,260 ±8 | 2.35 ±0.20 | 0.**77** | 0.03 |
| Collaborative | 18,660 ±7 | 2.27 ±0.19 | 0.**44** | 0.04 |
| Single-authored | 2,311±892 | 2.87 ±0.43 | 0.**32** | 0.05 |

The results suggest that the power-law distribution is a plausible one for the three distributions of citations under analysis.

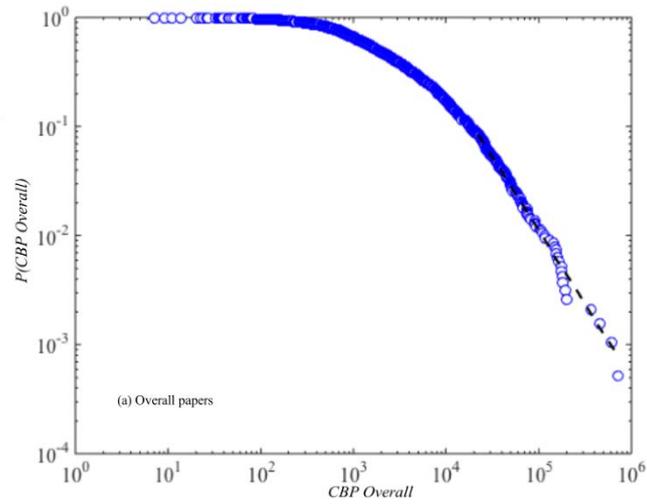

(a) Overall papers

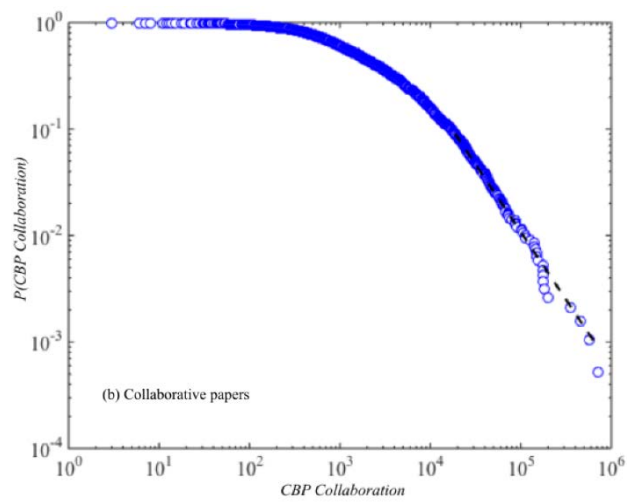

(b) Collaborative papers

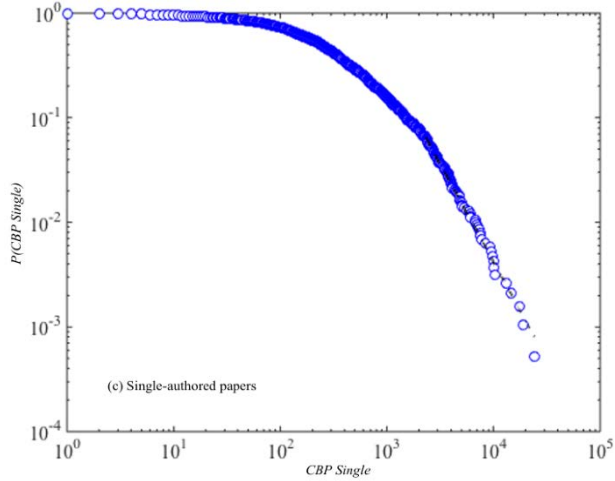

**Fig. 1** – Complementary cumulative probability distribution for the power law model of (a) overall papers, (b) collaborative papers and (c) single-author papers

*Comparing the power-law with alternative distributions*

To test the hypothesis for the power-law distribution we compared the power-law model to log normal, exponential and power-law with exponential cut-off as competing distributions by comparing loglikelihood ratios (LR). Positive values of LR indicate that the power-law model is favored over the alternative (Clauset et al., 2009). Table 4 shows the results of comparing the power-law hypothesis to other competing distributions. A power-law with exponential cut-off better fits these distributions than a pure power-law. This result supports hypothesis 1. A power-law with an exponential cut-off is a degenerate form of a power-law (Katz, 2015). While a pure power-law is scale-invariant from xmin to the end of the distribution a power-law with exponential cut-off is only scale-invariant from xmin to the point at the far right hand side of the distribution where the exponential decay begins dominates the power-law. The scale-invariant region can be many orders of magnitude in size. Some people think that the exponential cut-off of a power-law is due to the finite size of the data set, but recently it has been shown that it might also be an effect of finite observation time (Yamasaki et al., 2006). Moreover, models show that the probability distribution tends to evolve from exponential to a power-law with exponential cut-off to a pure power-law given enough time.

**Table 4.** Test of power-law behavior in the three datasets

| Dataset | Power-law | Log-normal | | exponential | | Power-law + cut-off | | Support for power-law |
|---|---|---|---|---|---|---|---|---|
| | *p* | LR | *p* | LR | *p* | LR | *p* | |
| Overall | 0.**77** | -0.32 | 0.30 | 39.51 | **0.00** | -1818 | **0.05** | With cut-off |
| Collaboration | 0.**44** | -0.77 | 0.22 | 2.83 | **0.00** | -1955 | **0.03** | With cut-off |
| Single-authored | 0.**32** | -0.46 | 0.26 | -1032 | **0.08** | -1096 | **0.04** | With cut-off |

**Power-law regression of collaboration predicting citation-based performance**

Table 5 shows the results of the analysis of the power-law correlations. The power-law correlation between the overall number of articles and the citation-based performance is highly significant $t(1, 31)= 9.824, p < 0.001$. The power-law correlation between collaborative papers and citation-based performance is statistical significant $t(1, 31)= 12.22, p < 0.001$. The power-law correlation between single-authored papers and citation-based performance is statistical significant $t(1, 31)= 5.02, p < 0.001$. We aggregated the years 2005 to 2006, 2006 to 2007 and the three years together and ran the power-law regression and we found out that the exponent of the correlation remain the same.

**Table 5**. Values of the exponents for the power-law correlations.

| Year | Overall | | | Collaboration | | | Single | | |
|---|---|---|---|---|---|---|---|---|---|
| | *Alpha* | *SD* | $R^2$ | *Alpha* | *SD* | $R^2$ | *Alpha* | *SD* | $R^2$ |
| 2005 | 1.19 | 0.08 | 0.88 | 1.20 | 0.07 | 0.91 | 0.91 | 0.12 | 0.66 |
| 2006 | 1.18 | 0.08 | 0.87 | 1.19 | 0.07 | 0.90 | 0.84 | 0.11 | 0.66 |
| 2007 | 1.19 | 0.08 | 0.87 | 1.20 | 0.07 | 0.90 | 0.84 | 0.11 | 0.66 |
| 2005-2006 | 1.19 | 0.08 | 0.88 | 1.20 | 0.07 | 0.91 | 0.87 | 0.11 | 0.67 |
| 2006-2007 | 1.18 | 0.08 | 0.87 | 1.20 | 0.07 | 0.90 | 0.84 | 0.11 | 0.67 |
| 2005-2007 | 1.19 | 0.08 | 0.87 | 1.20 | 0.07 | 0.91 | 0.86 | 0.11 | 0.67 |
| **Median** | **1.19** | **0.08** | **0.87** | **1.20** | **0.07** | **0.91** | **0.85** | **0.11** | **0.67** |

The results of aggregating two and three years in the analysis suggest that the exponent for the relationship between size and CBP to overall scientific output of natural sciences is about 1.19, the exponent for collaborative papers is 1.20 and 0.85 for single-authored papers. This means that no differences appear in the Mathew Effect by aggregating more years to the analysis.

Figure 2-A shows the results of the power-law correlation between citation-based performance and overall 2005-2007 papers. Figures 2B and 2C show the results of the power-law correlation between citation-based performance and collaboration and single-authored articles.

Table 5 shows the values for the scaling relationship of the three datasets studied. The result shows that the Mathew Effect (Merton, 1968, 1988) is bigger for collaborative articles. The median exponent is $1.20 \pm 0.7$. The exponent $> 1$ shows that citations grow faster than collaboration. This result supports the hypothesis 2.

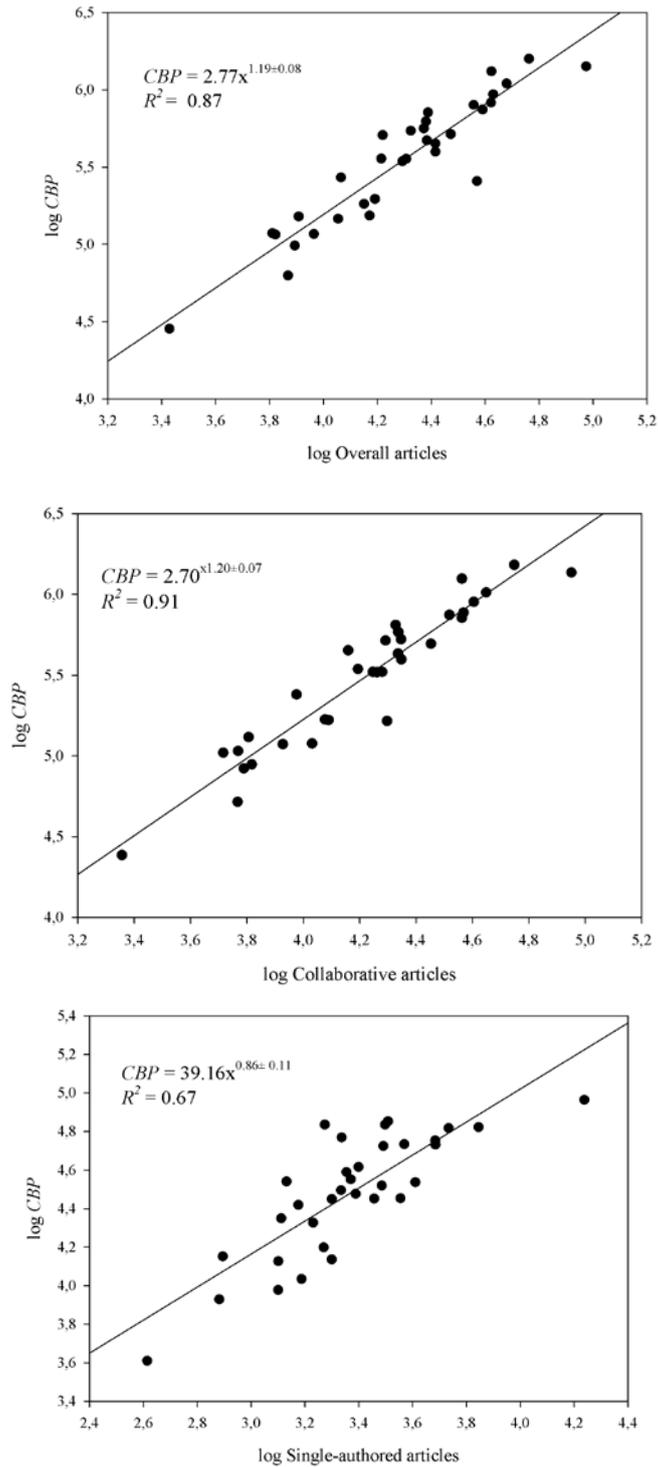

**Fig. 2** – Power law relationship between CBP and field sizes for (a) overall papers, (b) collaborative papers and (c) single-authored papers

The exponent for the relationship between citations and non-collaborative papers is 0.85 ± 0.11. An exponent < 1 suggests that single-authored papers exhibit an inverse Matthew effect. In other words, for a doubling of size the number of citations increases less than twice and in this case 1.80 times. A possible reason that the exponent for the single-authored papers is <

1.0 is that maybe a more interdisciplinary approach is required in large diverse fields to have greater impact.

The number of citations expected by collaborative papers increases $2^{1.20}$ or 2.30 times when the number of collaborative papers published in a subfield of the natural sciences doubles.. The Matthew effect is stronger for collaborative than for non-collaborative papers. In fact non-collaborative papers exhibit an inverse Mathew effect indicative of a cumulative disadvantage as Katz (2006b) suggested. The impact of single author articles decreases with larger subfields. According to these result, hypothesis 3 is sustained.

## Discussion and conclusions

In the present study we found three scaling correlations. Scaling relationships cannot be captured by traditional scientometrics indicators based on population averages e.g. citations per paper cannot capture the scaling correlation between impacts of collaborative and non-collaborative articles.

The exponent of the scaling correlation between citations and the number of collaborative/non-collaborative papers published by the scientific community of a research field is independent of the size of the system. It provides decision makers with a measure of the average expected impact for a research field given a known number of collaborative or non-collaborative papers. The measured citation-based impact for each natural science subfield can be compared against an expected impact providing a unique scale independent measure of performance that can be compared across subfields. For example, the scaling exponent for the correlation between citations and collaboration for the subfields studied was of $2^{1.20 \pm 0.07}$. This means that citation are expected to increase 2.30 times when a subfield doubles the number of papers it publishes through collaboration.

This result suggests that the citation-based performance of collaborative articles of a science system is greater than sole authorship. A science system can increase the impact of its knowledge by encouraging collaboration over no collaboration. Collaboration is a positive strategy for policy makers to foster greater impact of science systems.

## Possible new research questions

The present study suggests new research questions with policy implicatoins such as: Does international collaboration show a greater or lower Matthew effect than domestic collaboration? Such questions could be examined at the macro (regions, countries), meso (institutions, faculties, departaments, journals), and mico (researchers) levels. What is the impact of self-citations? Is the scaling exponent always < 1.0 for non-collaborative papers or does it does start out > 1.0 and become < 1.0 as time progresses? How is the magnitude of the scaling exponent for collaborative papers effected by the type of collaboration (international, domestic, inter-institutional, intra-institutional)?

## Acknowledgements

The authors thank Aaron Clauset for kindly answering our questions on the application of his procedure to test hypothesis on power-law distributions. To Professor Colin Gillespie for helping with us with the code of his computational program to analyze heavy tail distributions. To Professor Manuel Estay for helping with the Transact SQL queries, and linking the programs R and Matlab.